%% file: root.tex
\documentclass[letterpaper, 10 pt, conference]{ieeeconf}

\input{Packages_NewCommands}

\bibliography{references,dfk-refs, ref-mustafa}

\IEEEoverridecommandlockouts
\title{\LARGE \bf Scenario-Game ADMM: A Parallelized Scenario-Based Solver for Stochastic Noncooperative Games
}

\author{Jingqi Li$^1$, Chih-Yuan Chiu$^1$, Lasse Peters$^2$, Fernando Palafox$^{*3}$, Mustafa Karabag$^{*3}$, \\ Javier Alonso-Mora$^2$, Somayeh Sojoudi$^1$, Claire Tomlin$^1$, and David Fridovich-Keil$^3$ \thanks{$^*$Equal contribution. The authors are with: $^1$University of California, Berkeley, $^2$Delft University of Technology, $^3$University of Texas, Austin. Correspondence to \href{mailto:jingqili@berkeley.edu}{\tt jingqili@berkeley.edu}.}
}

\begin{document}

\maketitle

\pagestyle{plain}




\input{0_Abstract}

\input{1_Introduction}

\input{2_Related_Work}

\input{3_Preliminaries}
\input{4_Problem_Formulation}

\input{5_Methods}

\input{6_Results}
\input{7_Conclusion_Future_Work}
\printbibliography

\input{A_Appendix}

\input{A_Appendix_proof}






\end{document}

%% file: Packages_NewCommands.tex
\usepackage{graphicx}
\usepackage{adjustbox}
\usepackage{amssymb}
\usepackage{amsmath}
\usepackage{mathtools}

\usepackage[acronyms, shortcuts]{glossaries}

\usepackage{wrapfig}
\usepackage[labelfont=bf,figurename=Fig.,font=small]{caption}
\usepackage{subcaption}
\usepackage{lipsum}

\usepackage{siunitx}
\usepackage{xcolor}
\usepackage{hyperref}
\usepackage[ruled,vlined,linesnumbered]{algorithm2e}
\usepackage[noend]{algpseudocode}
\usepackage{ifthen}
\usepackage{dsfont}
\usepackage[natbib=true,backend=biber,style=numeric-comp,sorting=none,maxbibnames=99]{biblatex}
\usepackage{balance}
\usepackage{dirtytalk}
\usepackage{afterpage}

\usepackage{times}

\usepackage[capitalize,noabbrev,nameinlink]{cleveref}
\crefformat{equation}{(#2#1#3)}
\Crefname{algocfline}{Algorithm}{Algorithms}
\Crefname{algocf}{line}{lines}

\allowdisplaybreaks

\newcommand{\E}{\mathbb{E}}

\newcommand{\Prob}{\mathbb{P}}

\newcommand{\ra}{\rightarrow}
\newcommand{\R}{\mathbb{R}}

\newcommand{\mbb}{\mathbb}

\newcommand{\mc}{\mathcal}

\newcommand{\numplayers}{N}

\newcommand{\prob}[1]{\mbb{P}_\params\left(#1\right)}

\newcommand{\player}[1]{\textnormal{\textbf{P}$#1$}}

\newcommand{\numscenarios}{S}
\newcommand{\var}{x}
\newcommand{\vars}{\mathbf{\var}}
\newcommand{\aux}{w}
\newcommand{\auxs}{\mathbf{\aux}}
\newcommand{\cost}{f}
\newcommand{\constraint}{h}
\newcommand{\al}{\mc{L}}
\newcommand{\lm}{\lambda}
\newcommand{\lms}{\boldsymbol{\lambda}}
\newcommand{\pen}{\rho}

\newcommand{\varsdim}{{Nn}}
\newcommand{\zvar}{\mathbf{z}}

\newtheorem{remark}{Remark}
\newtheorem{assumption}{Assumption}
\newtheorem{proposition}{Proposition}
\newtheorem{definition}{Definition}
\newtheorem{lemma}{Lemma}
\newtheorem{theorem}{Theorem}

\newcommand{\err}{\delta}
\newcommand{\params}{\theta}
\newcommand{\paramdim}{d}
\newcommand{\paramdist}{p_\params}

\newcommand{\chance}{\epsilon}

\newacronym{admm}{ADMM}{alternating direction method of multipliers}
\newacronym{qp}{QP}{quadratic program}
\newacronym{gne}{GNE}{generalized Nash equilibrium}
\newacronym{mcp}{MCP}{mixed complementarity program}
\newacronym{licq}{LICQ}{linear independence constraint qualification}
\newacronym{clf}{CLF}{control Lyapunov function}
\newacronym{cbf}{CBF}{control barrier function}
\newacronym{ioc}{IOC}{inverse optimal control}
\newacronym{lqr}{LQR}{linear-quadratic regulator}
\newacronym{kkt}{KKT}{Karush–Kuhn–Tucker}
\newacronym{irl}{IRL}{inverse reinforcement learning}
\newacronym{mle}{MLE}{maximum likelihood estimation}
\newacronym{hji}{HJI}{Hamilton-Jacobi-Isaacs}
\newacronym[longplural=open-loop Nash equilibria,plural=OLNE]{olne}{OLNE}{open-loop Nash equilibrium}
\newacronym[longplural={partially observable Markov decision processes}]{pomdp}{POMDP}{partially observable Markov decision process}
\newacronym{svo}{SVO}{social value orientation}
\newacronym{ukf}{UKF}{unscented Kalman filter}
\newacronym{ibr}{IBR}{iterated best response}
\newacronym{awgn}{AWGN}{additive white Gaussian noise}
\newacronym{iqr}{IQR}{interquartile range}

\newcommand{\cf}{c.f.\@\xspace}

\newcommand{\jth}{$j^{\textnormal{th}}$\xspace}

%% file: 0_Abstract.tex
\begin{abstract}
Decision-making in multi-player games can be extremely challenging, particularly under uncertainty. In this work, we propose a new sample-based approximation to a class of stochastic, general-sum, pure Nash games, where each player has an expected-value objective and a set of chance constraints. This new approximation scheme inherits the accuracy of objective approximation from the established sample average approximation (SAA) method and enjoys a feasibility guarantee derived from the scenario optimization literature. 
We characterize the sample complexity of this new game-theoretic approximation scheme, and observe that high accuracy usually requires a large number of samples, which results in a large number of sampled constraints. To accommodate this, we decompose the approximated game into a set of smaller games with few constraints for each sampled scenario, and propose a decentralized, consensus-based ADMM algorithm to efficiently compute a generalized Nash equilibrium (GNE) of the approximated game. We prove the convergence of our algorithm to a GNE and empirically demonstrate superior performance relative to a recent baseline algorithm based on ADMM and interior point method.
\end{abstract}

%% file: 1_Introduction.tex
\section{Introduction}
\label{sec: Introduction}


Stochastic game theory \cite{shapley1953stochastic} provides a principled mathematical foundation for modeling interactions between multiple self-interested players in uncertain environments, and has applications in traffic control \cite{elhenawy2015intersection}, multi-robot coordination \cite{yan2013survey}, and human-robot interaction \cite{li2019differential}. In this framework, each player selects actions to optimize their own objective, obey a set of constraints, and reason about the strategic response of other players. Uncertainty in the players' potentially conflicting objectives and coupled constraints makes these problems extremely challenging to solve. 

Classical results in stochastic games are often derived under strong assumptions regarding the problem structure and the distribution of the underlying random process. 
In the class of linear quadratic Gaussian games, necessary and sufficient conditions for the existence of Nash equilibria are characterized in \cite{basar1998DynamicNoncooperativeGameTheory}. It is also shown that in $N$-player noncooperative stochastic games, the convexity of player-specific objectives and convex, compact strategy sets are sufficient for the existence of the Nash equilibria \cite{lei2022stochastic}. 
However, for general stochastic games, it is NP-hard to determine the existence of Nash equilibria \cite{conitzer2008new}. Moreover, computing a Nash equilibrium can also be a hard problem \cite{daskalakis2009complexity}, partially due to the complexity of solving the nonlinear equations induced by the Nash equilibrium condition.



Several recent efforts provide computationally efficient, approximate solutions to stochastic games with coupled constraints. Two lines of work provide high probability guarantees of both optimality and feasibility. The first \cite{xu2010sample,xu2013stochastic,peng2021stochastic} approximates the players' expected value objectives and constraints with sample average approximations. The second \cite{paccagnan2019scenario,fele2019probabilistic,fabiani2022probabilistic} follows the idea of scenario programming and approximates the objectives and constraints using worst-case samples. The former approach enjoys low sample complexity under certain distributional assumptions \cite{hu2020sample}. However, when the sample size is finite, this method may lead to situations in which the optimal solution is infeasible for the original chance constraint. 
The latter technique does not require strong distributional assumptions and returns conservative feasible solutions with high probability, but may require a large number of samples. 
In our work, we combine the benefits of the two approaches such that we obtain accurate approximations for the objectives and maintain the high probability feasibility guarantee. 

\looseness=-1
Our contributions are threefold: (1) We first propose a new sample-based approximation to the constrained stochastic game problem. In this framework, we approximate the expected objectives using a sample average approximation and ensure the feasibility of the original chance constraints by considering a large number of sampled constraints. We validate this scenario-game approximation by characterizing its sample complexity, and we show how the sample complexity can be improved by using problem structure. (2) 
To overcome the computational burden induced by the sampled constraints,
we decompose the approximated game into smaller games with few constraints per scenario, and propose a decentralized ADMM algorithm to compute the joint Nash equilibrium solution in parallel. (3) We prove the convergence of our method to a generalized Nash equilibrium of the approximated constrained game. Empirical results show that our method can handle a large number of constraints with faster convergence than a state-of-the-art baseline. 

%% file: 2_Related_Work.tex
\section{Related Work}
\label{sec: Related Work}



\subsection{Stochastic Games}
\label{subsec: Game Theory}

Originally due to \citet{shapley1953stochastic}, the field of stochastic game theory has expanded to model uncertainties in players' objectives \cite{harsanyi1967games}, constraints \cite{charnes1973prior}, and in the case of dynamic games, underlying state dynamics \cite{fink1964equilibrium,parthasarathy1989existence,nowak2003new}. 
Exact generalized Nash equilibrium solutions to stochastic constrained games can be obtained by solving their equivalent stochastic variational inequality problems \cite{facchinei2003finite,peng2021stochastic}. Under an appropriate constraint qualification, the well-known \ac{kkt} conditions must be satisfied for all players at a generalized Nash equilibrium \cite{facchinei2003finite, laine2021computation}. 
We focus on games with monotone objective pseudogradients \cite{pavel2019distributed} and convex constraints, where solutions can be found in polynomial time \cite{kinderlehrer2000introduction}.




\subsection{ADMM for Games}
\label{subsec: ADMM for Games}

We are ultimately interested in decentralized methods \cite{chen2022decentralized,scutari2010convex} for identifying generalized Nash equilibria, because they can often exploit computational parallelism for efficiency gains compared with centralized method. In particular, the \ac{admm} \cite{gabay1976dual, boyd2011distributed} is an appealing approach for efficient decentralized computation.
The ADMM enjoys convergence guarantees for convex problems \cite{glowinski1975approximation, boyd2004convex}, convex-concave saddle point problems \cite{bredies2015preconditioned,karabag2022alternating} and monotone variational inequality problems \cite{he2002new,yang2022solving}. 
Recent work \cite{yang2022solving} has adopted an interior-point method to ensure constraint feasibility, thereby outperforming projection-based consensus ADMM methods \cite{korpelevich1976extragradient,nemirovski2004prox,diakonikolas2020halpern}.

\looseness=-1
Our algorithm differs from prior work \cite{borgens2021admm,salehisadaghiani2017distributed,le2020parametrized} in that we decompose the objective and constraints \emph{over scenarios}. For each scenario, we solve an \(N\)-player game with relatively few constraints, and then synchronize across scenarios via \ac{admm}. Unlike prior methods, we do not require constraint projection or an interior-point method in the consensus step. Moreover, we can handle nonlinear coupled constraints, while prior works \cite{liang2017distributed,pavel2019distributed} consider affine constraints. 





\subsection{Approximation Methods for Stochastic Optimization}
\label{subsec: Scenario Methods}


The sample average approximation (SAA) method \cite{vogel1988stability} is a well-known technique for solving stochastic optimization problems via Monte Carlo simulation \cite{homem2014monte}. This method approximates the objectives and constraints of the original problem using sample averages, and has been shown to be able to recover original optimal solutions, as the sample size grows to infinity \cite{xu2010sample, xu2013stochastic, peng2021stochastic}. Another approach for approximating the stochastic optimization problem is the scenario optimization approach \cite{dembo1991scenario, calafiore2006scenario}, where the original chance constraints are replaced with a large number of sampled constraints \cite{calafiore2005uncertain}. This method has been extensively studied, and subsequent work has characterized its sample complexity and feasibility guarantees \cite{campi2008exact}. Moreover, it is recently extended to constrained variational inequality problems~\cite{paccagnan2019scenario}. Our approach approximates the expected value objective by a sample average, and replaces the chance constraint with a large number of sampled constraints. 





%% file: 3_Preliminaries.tex
\section{Preliminaries}
\label{sec: Preliminaries}



We begin by introducing a deterministic, general-sum static game played among $\numplayers$ players. Concretely, each player $i$ (\player{i}) seeks to solve a problem of the form:
\begin{subequations}\label{eqn:deterministic-game}
\begin{align} \label{Eqn: Game, deterministic, Objective}
    \var_i^* \in \arg \min_{\var_i}~&\cost_i(\vars)\\ \label{Eqn: Game, deterministic, Constraint}
    \mathrm{s.t.}~&\constraint_i(\vars) \le 0\,,
\end{align}
\end{subequations}
where $x_i\in \mathcal{X}_i\subseteq \R^n$, for each $i\in [N]:=\{1,2,\dots,N\}$, $\mathcal{X}_i$ is the domain of $x_i$ and $\vars := (\var_1, \dots, \var_\numplayers) \in  \R^\varsdim$. Let the joint decision space be denoted by $\mathcal{X}:=\mathcal{X}_1\times\dots\times \mathcal{X}_N$, and let each player $i$'s constraint be denoted by $h_i(\vars):\mathcal{X}\to\mathbb{R}^\ell$.
Observe that players' problems are \emph{coupled}, both in the objectives and the constraints. 
We are interested in finding unilaterally optimal strategies for all players in this setting, i.e., the generalized Nash equilibria.
\begin{definition}[\cite{facchinei2010generalized}]
    A point $\vars^* \in \R^\varsdim$ is a \textit{generalized Nash equilibrium} (GNE) if for all $i\in[N]$, $h_i(\vars^*)\le 0$, and $f_i(\var_i, \vars_{-i}^*) \ge f_i(\vars^*)$, for each $\var_i$ satisfying $h_i(\var_i, \vars_{-i}^*) \le 0$.
\end{definition}

%% file: 4_Problem_Formulation.tex
\section{Scenario Game Problem}\label{sec:problem formulation}

In this work, we focus our attention on constrained \emph{stochastic} general-sum games, in which both the objective and constraints are subject to uncertainty and parameterized by the random vector $\params$, i.e. $\cost_i(\vars; \params)$ and $\constraint_i(\vars; \params)$.
Let the random vector of parameters $\params \in \Theta \subseteq \mbb{R}^\paramdim$ 
be drawn from a probability distribution $\paramdist$ that is unknown to all players. 
We denote player \emph{i}'s decision problem as:
\begin{subequations}
\label{eqn:stochastic-game}
\begin{align}
    \var_i^* \in \arg \min_{\var_i}~&\E\left[\cost_i(\vars; \params)\right]\\
    \mathrm{s.t.}~&\Prob_\params \big( \constraint_i(\vars; \params) \le 0 \big) \ge 1 - \chance\,.
\end{align}
\end{subequations}
Note that we have replaced \player{i}'s objective with its expectation under distribution $\paramdist$, and likewise we have replaced the deterministic constraint $\constraint_i(\vars;\params) \le 0$ with the chance constraint $\prob{\constraint_i(\vars;\params) \le 0} \ge 1 - \chance$, with $\chance \in (0, 1)$ as the probability of failure.
In full generality---i.e., without making further assumptions about the distribution $\paramdist$, such as normality---it is intractable to find a generalized Nash equilibrium for \cref{eqn:stochastic-game}.
In the sequel, we will construct a sampled approximation to \cref{eqn:stochastic-game} which is amenable to both theoretical complexity analysis and efficient, parallel implementation.

Drawing upon ideas developed in the stochastic optimization \cite{dembo1991scenario,calafiore2005uncertain} and model predictive control \cite{calafiore2006scenario,calafiore2012robust,campi2018general} communities, we approximate the stochastic game \cref{eqn:stochastic-game} with the following deterministic problem:
\begin{subequations}
\label{eqn:scenario-game}
\begin{align}
    \var_i^* \in \arg \min_{\var_i}~&\frac{1}{\numscenarios}\sum_{j = 1}^\numscenarios \cost_i(\vars; \params^j)\\
    \mathrm{s.t.}~&\constraint_i(\vars; \params^j) \le 0,~\forall j \in \{1, \dots, \numscenarios\}\,,
\end{align}
\end{subequations}
in which each so-called \emph{scenario} $\params^j$ is sampled independently from the probability distribution $\paramdist$.
In \cref{eqn:scenario-game}, we have replaced the expected value of the objective from \cref{eqn:stochastic-game} with its empirical mean, and enforced the original constraint in \cref{eqn:deterministic-game} for all of the scenarios $\{\params^j\}_{j=1}^\numscenarios$. We propose to compute the generalized Nash equilibrium of \eqref{eqn:scenario-game}, {which always exists if the following assumption holds true \cite{flaam1993paths}}.



\looseness=-1
\begin{assumption}\label{assumption: ADMM convergence}
For each player $i\in[N]$, the constraint $h_i(\vars;\params)$ is convex in $\vars$ and satisfies Slater's condition \cite{boyd2004convex}. The objective function of each player is upper bounded, i.e. $\sup_{\params \in \Theta,\vars\in \mathcal{X}}\|f_i(\vars;\params)\|_\infty\le D$, for some finite $D\in\mathbb{R}$. 
The pseudogradient $F(\vars;\theta):=[\nabla_{x_i} f_i(\vars;\theta)]_{i=1}^N$, where $\nabla_{x_i} f_i(\vars;\theta)$ denotes the gradient of $f_i$ with respect to $x_i$, is a continuous and monotone operator of $\vars$, i.e., $(\vars-\mathbf{y})^\top(F(\vars;\theta) - F(\mathbf{y};\theta))\ge 0,\forall \vars,\mathbf{y}\in\mathbb{R}^{Nn}$.
\hspace{1.2cm}$\square$
\end{assumption}

Assumption~\ref{assumption: ADMM convergence} implies that the objective of each player is convex with respect to its own decision variable, a standard assumption in variational inequality problems \cite{yang2022solving}. It is shown in \cite{jiang2022generalized} that a convex-concave saddle point problem can be reformulated to satisfy Assumption \ref{assumption: ADMM convergence}. Note also that Assumption~\ref{assumption: ADMM convergence} allows \emph{non}convex objectives for each player. An example is a two-player game, with the objectives $f_1(x_1,x_2) = x_1 - x_2^2$ and $f_2(x_1,x_2) = x_2 - x_1^2$.

\textbf{Running Example:} We consider a simplified spacecraft rendezvous problem, where two spacecraft approach each other at a predefined rendezvous point in space. We model this problem as a two player general-sum game with a planning horizon $T$. At time $t\in\{0,1,\dots,T\}$, each spacecraft has a state vector 
$\xi_i(t)=[\xi_i^x(t),\xi_i^{v_x}(t), \xi_i^y(t), \xi_i^{v_y}(t)] \in \mathbb{R}^{4}$, where $[\xi_i^x(t),\xi_i^y(t)]$ is the position of the spacecraft in the rendezvous hyperplane and $[\xi_i^{v_x},\xi_i^{v_y}]$ is the velocity vector. It also has a control vector $u_i(t)=[u_i^x(t), u_i^y(t)]\in\mathbb{R}^2$ representing the x- and y-axis acceleration. The dynamics of each spacecraft is approximated as a double integrator for simplicity~\cite{harris2014minimum},
\begin{equation}
    \xi_i(t+1) =\underbrace{\left[\begin{matrix}  1 & \Delta t & 0 & 0 \\ 0 & 1 & 0 & 0 \\ 0 & 0 & 1 & \Delta t \\ 0 & 0 & 0 & 1  \end{matrix} \right]}_{A} \xi_i(t) + \underbrace{\begin{bmatrix} \frac{1}{2}\Delta t^2 & 0\\ \Delta t & 0 \\ 0 & \frac{1}{2}\Delta t^2 \\ 0 & \Delta t \end{bmatrix}}_{B}u_i(t)
\end{equation}
where $\Delta t>0$ is the time discretization constant. 
We assume the initial state $\xi_i(0)$ is drawn from a known distribution $p_{\xi}$. We concatenate all the random parameters into a vector $\params \in\mathbb{R}^d$, and assume it follows a distribution $\theta\sim p_\params$. As such, the general-sum game that each player considers can be summarized as follows,
\begin{equation}\label{eq: running example}
\begin{aligned}
    \min_{\{u_i(t)\}_{t=0}^{T-1}} &\frac{1}{T}\sum_{t=0}^T \mathbb{E}_\theta\left[\frac{1}{2} \xi_i(t)^\top Q_i^\theta\xi_i(t) + \frac{1}{2} u_i(t)^\top u_i(t) \right] \\
    \textrm{s.t. }
    & \mathbb{P}_\theta\Big(\left[\begin{smallmatrix}
        \xi_1^x(t) - \xi_2^x(t)\\ \xi_1^y(t) - \xi_2^y(t)
    \end{smallmatrix}\right]-b_i^\params \le 0, \|u_i(t)\|_\infty\le 1, 
     \\ &\ \ \ \ \  \left\|\left[\begin{smallmatrix}
        \xi_1^x(t) - \xi_2^x(t)\\ \xi_1^y(t) - \xi_2^y(t)
\end{smallmatrix}\right]\right\|_2^2\le 1, 
     \forall t\in[T]\Big)\ge 0.95
\end{aligned}
\end{equation}
where $\xi_i(t+1) = A\xi_i(t) + Bu_i(t)$, $\forall i\in\{1,2\}$, $\forall t\in\{0,1,\dots,T-1\}$, and $b_i^\params$ parameterizes an inequality chance constraint ensuring no hard contact between two spacecraft with high probability. Note that each player's feasible set depends upon the decisions of the other player. Hence, this is a \emph{generalized} Nash equilibrium problem.
In the following sections, we will discuss how many samples are required such that we can approximate \eqref{eq: running example} well using \eqref{eqn:scenario-game}, and develop an efficient method for computing a generalized Nash equilibrium of the sample-approximated game.



%% file: 5_Methods.tex
\section{Sample Complexity of Scenario Games}
\label{sec: Methods}
One of the appealing aspects of scenario programming \cite{campi2008exact,paccagnan2019scenario} is its generality with respect to the distribution of parameter vector $\theta$. Indeed, one can establish probabilistic guarantees on the feasibility of the original chance constraint without strong assumptions that $p_{\theta}$ be, e.g. sub-Gaussian or sub-exponential. We extend this result to the scenario game problem, and characterize sample complexity as follows:
\begin{proposition}\label{prop: sample complexity}
    Consider $\epsilon,\delta\in (0,1)$ and $\tilde{\epsilon}> 0$. Let $\{\theta^j\}_{j=1}^S$ be i.i.d. samples of the random variable $\theta\sim p_\theta$. Let $S$ be the sample size. Define $\mathcal{H}_S := \{\vars\in\mathbb{R}^{Nn}: h_i(\vars;\params^j)\le 0,\forall i \in [N], j\in [S]\}$. Suppose that $\mathcal{H}_S$ is non-empty, then under Assumption~\ref{assumption: ADMM convergence}, the following statements hold true simultaneously for each player $i\in[N]$,
    \begin{enumerate}
        \item $\| \frac{1}{S}\sum_{j=1}^S f_i(\vars;\params^j) - \mathbb{E}_{\params}[f_i(\vars;\params)]\|\le \tilde{\epsilon}$, for all $x\in\mathcal{H}_S$
        \item $\mathbb{P}_\params(h_i(\vars;\params)\le 0)\ge 1-\epsilon $, for all $x\in\mathcal{H}_S$
    \end{enumerate}
    with probability at least $1-\delta$, where $\delta:=2Ne^{-\frac{S\tilde{\epsilon}^2}{4D^2}} +   \sum_{\ell = 0}^{Nn-1}\binom{S}{\ell}\epsilon^{\ell}(1-\epsilon)^{S - \ell}$.
\end{proposition}

\begin{proof}
    The proof can be found in the Appendix.
\end{proof}

{
Note that we have not made strong assumptions on the distribution $p_\params$; the bound can be improved if more prior knowledge about the problem structure and distribution $p_\params$ is available. 
For example, if each player's constraint $h_i(\vars;\params)\le 0$ only depends on its own decision variable $\var_i$, then the constraint $h_i(\vars;\params)$ can be simplified as $h_i(\var_i;\params)\le 0$, where the decision variable $\var_i\in \mathbb{R}^n$ has a lower dimension than the original decision variable $\vars\in\mathbb{R}^{Nn}$. This dimension reduction simplifies the sample complexity for approximating each constraint. By combining this simplification with the union bound, we can improve the sample complexity result of Proposition \ref{prop: sample complexity}, as shown in the following result.}
\begin{proposition}\label{prop: improved sample complexity}
    Under the same assumptions of Proposition~\ref{prop: sample complexity}, suppose that $\mathcal{H}_S$ is non-empty and each player's constraint $h_i(\vars;\params^j)\le 0$ only depends on $\var_i$, $\forall j\in[S]$. Then, the following statements hold true simultaneously for each player $i\in[N]$,
    \begin{enumerate}
        \item $\| \frac{1}{S}\sum_{j=1}^S f_i(\vars;\params^j) - \mathbb{E}_{\params}[f_i(\vars;\params)]\|\le \tilde{\epsilon}$, for all $x\in\mathcal{H}_S$
        \item $\mathbb{P}_\params(h_i(\vars;\params)\le 0)\ge 1-\epsilon $, for all $x\in\mathcal{H}_S$
    \end{enumerate}
    with probability at least $1-\delta$, where $\delta:=2Ne^{-\frac{S\tilde{\epsilon}^2}{4D^2}} +  N \sum_{\ell = 0}^{n-1}\binom{S}{\ell}\epsilon^{\ell}(1-\epsilon)^{S - \ell}$. 
\end{proposition}

\begin{proof}
    The proof can be found in the Appendix.
\end{proof}

The above characterization of sample complexity suggests that a sufficient number of samples leads to an accurate estimation of the objective and ensures the feasibility of the chance constraint with high probability. 
However, 
solving a constrained game with a large number of sampled constraints presents a significant computational challenge. This motivates the following algorithmic development.

    
        

        

\section{Scenario Games via Decentralized ADMM}\label{sec:scenario games via decentralized ADMM}

\begin{algorithm}[tbp]
\DontPrintSemicolon
\caption{Scenario-Game ADMM (SG-ADMM)}\label{alg:scenario-admm}
{
\small
    \textbf{Input:} Initialization $\{\auxs(0),\vars(0),\lms(0)\}$, convergence tolerance $\epsilon>0$.
    
    \For{$k=0,1,2,\cdots$}{
        \For{scenarios \(j=1, \dots, S\) in parallel\label{alg:parallel solving each scenario}}{$\begin{aligned}
            \aux_i^{j}(k+1) 
            \gets \arg\min_{\aux_i^j}~ & \mathcal{L}_i^j(\auxs^j,\vars(k),\lms_i(k))\\
            \textrm{s.t. }& h_i(\auxs^j;\params^j)\le 0
            \end{aligned}$
        }
        
        Update $\{\var_i(k+1)\}_{i=1}^N$:  $\forall i\in[N]$,
        $\var_i(k+1)\gets \frac{1}{S}\sum_{j=1}^S(\frac{1}{\rho}\lm_i^j(k) + \aux_i^j(k+1)) $

        Update $\{\lm_i^j(k+1)\}_{i=1,j=1}^{N,S}$: $\forall i\in[N]$, $j\in[S]$,
        $\lm_i^j(k+1)\gets \lm_i^j(k) + \rho (\aux_i^j(k+1)-\var_i(k+1))$
        
        \textbf{If} $\|\auxs(k+1)-M\vars(k)\|^2\le \epsilon$, \textbf{return} $\{\var_i(k+1)\}_{i=1}^N$
    }

}
\vspace{-0.25em}
\end{algorithm}

\subsection{Decentralized ADMM}

In the scenario game \cref{eqn:scenario-game}, both the objective and constraints involve significantly more terms than in \cref{eqn:deterministic-game}.
When $\numscenarios$ is large, therefore, it can be computationally demanding to find a generalized Nash equilibrium.
Therefore, we propose the following splitting method to enable parallel---and hence more efficient---computation of equilibrium solutions.
This technique is an analog of the well-known \ac{admm} algorithm tailored to generalized Nash equilibrium problems, and is summarized in \cref{alg:scenario-admm}.

In order to develop this technique, we shall begin by introducing auxiliary decision variables $\{\aux_i^j\}_{j=1}^\numscenarios$ for each player \player{i}, and employing the shorthand $\auxs_i := (\aux_i^1,\dots,\aux_i^S)$ for the decision variables of player $i$ across scenarios $j=1$ to $j=S$, $\auxs^j := (\aux_1^j,\dots,\aux_N^j)$ for the decision variables of players $i=1$ to $i=N$ in the $j$th scenario, and $\auxs:=(\auxs_1,\dots,\auxs_N)$ for all the decision variables. We will later use the same shorthand $(\lms_i, \lms^j,\lms)$ for Lagrange multipliers for the constraints \cref{eqn:scenario-consensus-constraint}:
\begin{subequations}\label{eqn:scenario-admm-1}
\begin{align} 
    \var_i^*, \auxs_i^* \in \arg \min_{\var_i, \auxs_i}~&\frac{1}{\numscenarios} \sum_{j = 1}^\numscenarios \cost_i(\auxs^j; \params^j)\\
    \label{eqn:scenario-inequality-constraint} \mathrm{s.t.}~&\constraint_i(\auxs^j; \params^j) \le 0,~\forall j \in \{1, \dots, \numscenarios\} \\
    \label{eqn:scenario-consensus-constraint}
    &\aux_i^j - \var_i = 0,~\hspace{0.35cm}\forall j \in \{1, \dots, \numscenarios\}.
\end{align}    
\end{subequations}

In \cref{eqn:scenario-admm-1}, \player{i} evaluates its objective and constraints for scenario $j$ using only the auxiliary variables $\auxs^j$.
However, in the end, each player must select a single decision variable; hence, we also enforce the consensus constraints in \cref{eqn:scenario-consensus-constraint}.
These constraints effectively \emph{couple} $\numscenarios$ games which would otherwise be entirely independent.
To facilitate such a decomposition, 
we construct a \emph{partial} augmented Lagrangian for each player, in which only \cref{eqn:scenario-consensus-constraint} have been dualized:
\begin{align}
\label{eqn:augmented-lagrangian}
    \al_i(\auxs,\vars,\lms_i) &:= \sum_{j = 1}^\numscenarios \al_i^j(\auxs^j,\vars,\lms_i), \\ \nonumber
    \al_i^j(\auxs^j,\vars,\lms_i) &:= \frac{\cost_i(\auxs^j; \params^j)}{\numscenarios} + \lm_i^{j\top} \err_i^j + \frac{\rho}{2} \|\err_i^j\|_2^2\,.
\end{align}
Here, $\err_i^j := \aux_i^j - \var_i$, and $\lm_i^j$ may be interpreted as an estimate of the Lagrange multiplier corresponding to the \jth instance of \cref{eqn:scenario-consensus-constraint} in \player{i}'s problem.
Thus equipped, we develop the key steps of \cref{alg:scenario-admm}, a decentralized technique for solving \cref{eqn:scenario-game} via \cref{eqn:scenario-admm-1}.
To do so, we re-express \cref{eqn:scenario-admm-1} in terms of the augmented Lagrangians \cref{eqn:augmented-lagrangian}:
\begin{subequations}
\label{eqn:scenario-admm-2}
\begin{align}
    \var_i^*,\auxs_i^* \in \arg \min_{\var_i, \auxs_i}~ &\al_i(\auxs, \vars,\lms_i)\\
    \mathrm{s.t.}~&\constraint_i(\auxs^j; \params^j) \le 0,\forall j \in \{1, \dots, \numscenarios\}\,.
\end{align}
\end{subequations}

\subsubsection{Solving for auxiliary variable, \texorpdfstring{$\auxs$}{\textbf{w}}}
\label{sec:solving-aux}

Holding $\vars$ and $\lms$ constant, each player's problem \cref{eqn:scenario-admm-2}
is convex in the decision variable $\auxs_i$ due to Assumption~\ref{assumption: ADMM convergence}.
Thus, we can be assured that any point $\auxs_i^*$ which satisfies the \ac{kkt} conditions for all players simultaneously is a generalized Nash equilibrium.
Such a point may be identified by, e.g., reformulating the joint \ac{kkt} conditions as a \ac{mcp} \cite{ferris2005mathematical} and invoking a standard solution method, e.g. PATH \cite{dirkse1995path}.


\begin{remark}
\label{remark:parallelizable}
This equilibrium problem may be separated into $\numscenarios$ independent problems, involving distinct variables $\{\auxs^j\}_{j=1}^\numscenarios$, objectives, and constraints.
Consequently, if parallel computation is available, these games may be solved in separate computational threads or on separate computer processors; therefore, \cref{alg:scenario-admm} may still operate efficiently and converge even when many scenarios are required, as shown in Theorem~\ref{thm:ADMM convergence}.
\end{remark}

\subsubsection{Solving for consensus variables, \texorpdfstring{$\vars$}{\textbf{x}}}
\label{sec:solving-consensus}

Holding $\auxs$ and $\lms$ fixed, player $i$'s problem \cref{eqn:scenario-admm-2} may be simplified to take the following form:
\begin{equation}
    \label{eqn:scenario-admm-3}
    \var_i = \arg \min_{\tilde \var_i}~ \sum_{j=1}^\numscenarios \left(\lm_i^{j\top} (\aux_i^j - \tilde \var_i) + \frac{\pen}{2}\|\aux_i^j - \tilde \var_i\|^2_2\right).
\end{equation}
Because $\pen > 0$, we readily identify the global solution to \cref{eqn:scenario-admm-3} for each player as
\begin{equation}
    \label{eqn:updating-consensus}
    \var_i \gets \frac{1}{\numscenarios} \sum_{j=1}^\numscenarios \left(\frac{1}{\pen} \lm_i^j + \aux_i^j\right)\,.
\end{equation}

\subsubsection{Updating dual variables, \texorpdfstring{$\lms$}{lambdas}}
\label{sec:updating-duals}

In order to choose new values of the dual variables which account for the solutions to the previous subproblems, we first examine player $i$'s vanishing gradient condition.
We find:
\begin{align}
    \label{eqn:dual-update-1}
    0&=  \partial_{\aux_i^j} (\al_i^j(\auxs^j,\vars,\lms_i) +  \mathbb{I}_{h_i(\auxs^j;\params^j)}(\auxs^j)) \\ =&  \dfrac{\nabla_{\aux_i^j} \cost_i(\auxs^j; \params^j)}{\numscenarios} + \partial_{\aux^{j}_i}\mathbb{I}_{ h_i(\auxs^j;\params^j)}(\auxs^j) + \lm_i^j + \pen (\aux_i^j - \var_i),\nonumber
\end{align}
where $\mathbb{I}_{h(\vars;\params)}(\vars):\mathbb{R}^{Nn}\to \{0,\infty\}$ and $\mathbb{I}_{h(\vars;\params)}(\vars)=0$ if and only if $h(\vars;\params)\le 0$.
Following well-established reasoning for augmented Lagrangian methods \cite[Ch. 17]{nocedal2006optimizationbook}, we recognize the latter two terms as the (unique) value of the Lagrange multiplier for the original constraint \cref{eqn:scenario-consensus-constraint} which satisfies the vanishing gradient optimality condition.
Therefore, we set:
\begin{equation}
    \label{eqn:dual-update-2}
    \lm_i^j \gets \lm_i^j + \pen (\aux_i^j - \var_i)\,.
\end{equation}
The above update rule is formalized in Algorithm~\ref{alg:scenario-admm}.


\subsection{Convergence of Scenario-Game ADMM}

In this section, we first characterize the optimality condition of the general-sum game problem. We then show that the special structure of the consensus constraint allows us to measure convergence by monitoring the residual of the consensus constraint. Building upon this result, we prove the convergence of Algorithm~\ref{alg:scenario-admm}.

Similar to standard, single-objective optimization problems, under an appropriate constraint qualification the KKT conditions must be satisfied at solutions to the variational inequality problem \cite{facchinei2003finite}. From the KKT conditions, an optimal solution $\zvar^*:=(\auxs^*,\vars^*,\lms^*)$ should satisfy the following conditions, 
\begin{equation}\left\{
    \begin{aligned}
        &(p-\aux_i^{j*})^\top ({\nabla_{\aux_i^{j*}} f_i(\auxs^{j*};\theta^j) +\partial_{\aux_i^{j*}}\mathbb{I}_{h_{i}(\auxs^{j*};\theta^j)}(\auxs^{j*})}\\ &  \ \ + \lms^*)\ge 0,\forall p\in\mathbb{R}^n,\forall i\in[N],j\in[S]\\ 
        & \auxs^* - M\vars=0\\
        & h_i(\auxs^{j*};\params^j)\le 0,\forall i\in[N], \forall j\in[S]
    \end{aligned}\right.\label{eqn: optimal condition for variational inequality}
\end{equation}
where we represent the consensus constraint \eqref{eqn:scenario-consensus-constraint} compactly as $\auxs-M\vars=0$ by introducing a constant matrix $M:=\mathbf{1}_N\otimes I_n$.
Let $F(\auxs):=[\nabla_{\aux_i^j} f_i(\auxs^j;\params^j)]_{i=1,j=1}^{N,S}$ and $H(\auxs): = [\partial_{\aux_i^j} \mathbb{I}_{h_i(\auxs^j;\params^j)}(\auxs^j)]_{i=1,j=1}^{N,S}$. We can also represent the above optimality condition as the variational inequality problem:
\begin{equation}\label{eq:VI optimality condition}
    (\zvar-\zvar^*)^\top Q(\zvar^*)\ge 0, \forall \zvar\in \mathbb{R}^{SNn}\times \mathbb{R}^{Nn}\times \mathbb{R}^{Nm}, 
\end{equation}
\begin{equation}
\begin{aligned}
    \zvar =& \begin{bmatrix} \auxs \\ \vars \\ \lms
    \end{bmatrix}, Q(z) = \begin{bmatrix} F(\auxs) + H(\auxs) + \lms \\ -M^\top \lms \\ \auxs-M \vars \end{bmatrix}.
\end{aligned}
\end{equation}

Observing that the $M$ matrix in the consensus constraint has full column rank, we see that it must have trivial null space. Consequently, we can show in the following lemma that an optimal solution is reached when the consensus constraint residual is zero.


\begin{lemma}\label{lem: consensus constraint residual}
Suppose $\auxs(k+1) - M \vars(k) = 0$, then $(\auxs(k+1),\vars(k+1), \lms(k+1))$ is an optimal solution to the VI problem \cref{eq:VI optimality condition}.
\end{lemma}
\begin{proof}
    The proof can be found in the Appendix.
\end{proof}

Building upon this result, we show in the following theorem that a Lyapunov function, defined by the Lagrange multiplier error and the consensus constraint's residual, is monotonically decreasing with each iteration of Algorithm~\ref{assumption: ADMM convergence}. 


\begin{figure}[t!]
    \centering
    \begin{subfigure}[b]{0.49\linewidth}
        \centering
        \includegraphics[width=\linewidth,trim = 10 20 10 0]{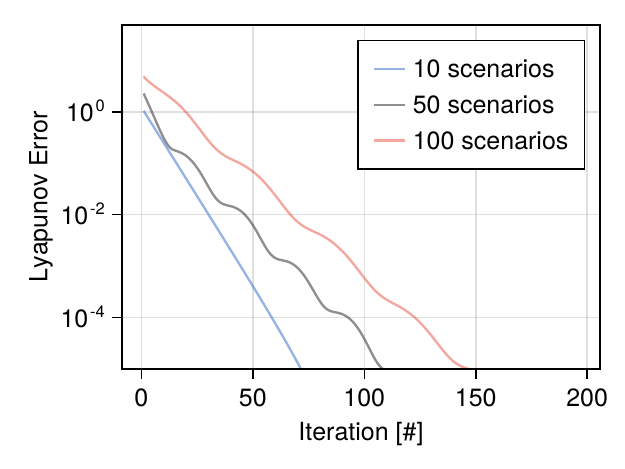}
        \caption{Lyapunov function}\label{subfig:lyapunov}
    \end{subfigure}
    \hfill
    \begin{subfigure}[b]{0.49\linewidth}
        \centering
        \includegraphics[width=\linewidth,trim = 10 20 10 0]{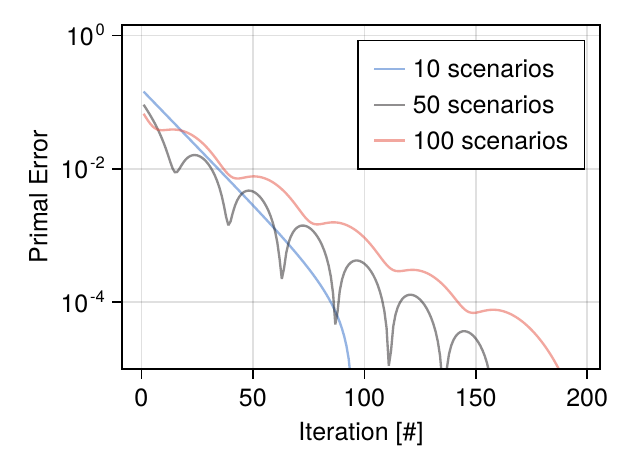}
        \caption{Primal residual}\label{subfig:primal residual}
    \end{subfigure}
    \caption{The convergence of Scenario-Game ADMM under different numbers of sampled scenarios in running example \eqref{eq: running example}. With only 10 samples, we have no binding constraint, and we converge exponentially fast. With 50 and 100 samples, we suffer binding constraints, and the primal residual $\rho\|M(\vars(k)-\vars^*)\|^2$ oscillates. However, the Lyapunov function, which is defined as the sum of primal residual and dual residual $\frac{1}{\rho}\|\lms(k)-\lms^*\|^2$, decays monotonically.}
    \vspace{-0.7cm}
    \label{fig:scenario-game ADMM convergence}
\end{figure}

\begin{figure}[t!]
    \centering
    \begin{subfigure}[b]{0.49\linewidth}
        \centering
        \includegraphics[width=\linewidth,trim = 10 20 10 0]{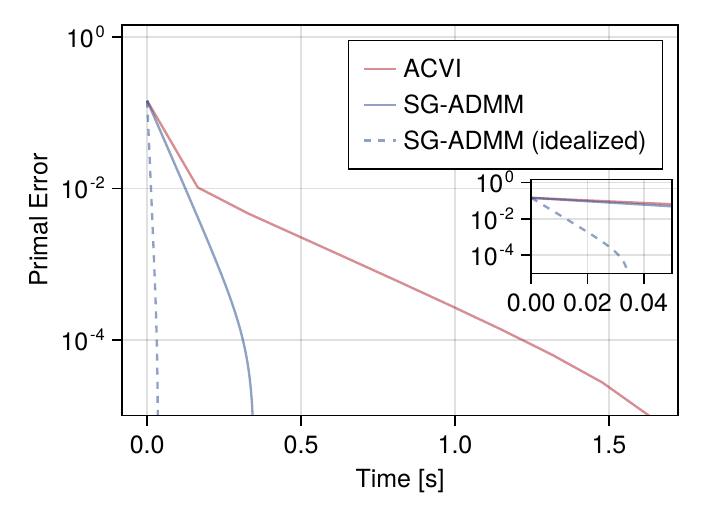}
        \caption{10 sampled scenarios, under linear dynamics \eqref{eq: running example}}\label{subfig:size 10}
    \end{subfigure}
    \hfill
    \begin{subfigure}[b]{0.49\linewidth}
        \centering
        \includegraphics[width=\linewidth,trim = 10 20 10 0]{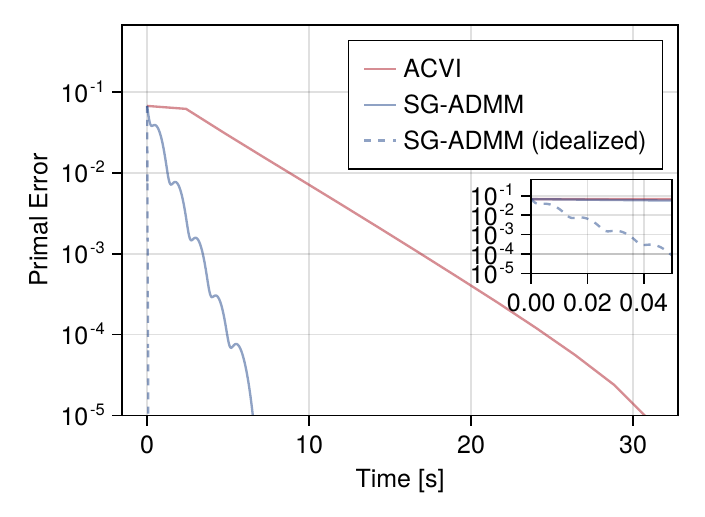}
        \caption{100 sampled scenarios, under linear dynamics \eqref{eq: running example}}\label{subfig:size 50}
    \end{subfigure}
        \begin{subfigure}[t]{0.49\linewidth}
        \centering
        \includegraphics[width=\linewidth,trim = 10 20 10 0]{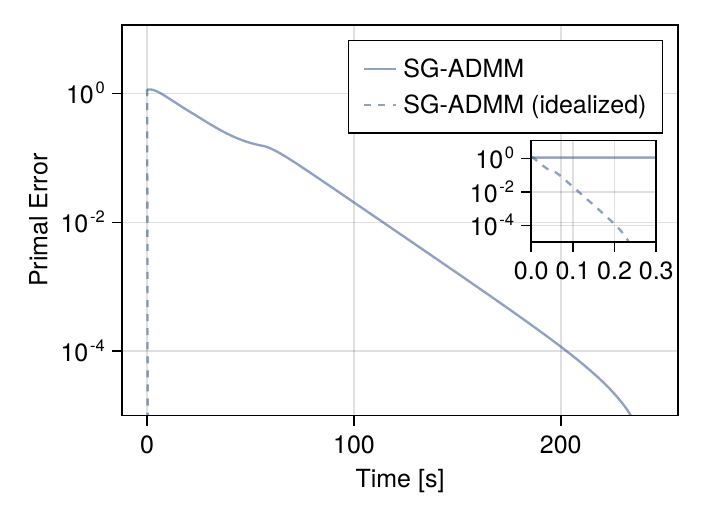}
        \caption{1000 sampled scenarios, under linear dynamics \eqref{eq: running example}}\label{subfig:size 100}
    \end{subfigure}
    \hfill
    \begin{subfigure}[t]{0.49\linewidth}
        \centering
        \includegraphics[width=\linewidth,trim = 10 20 10 0]{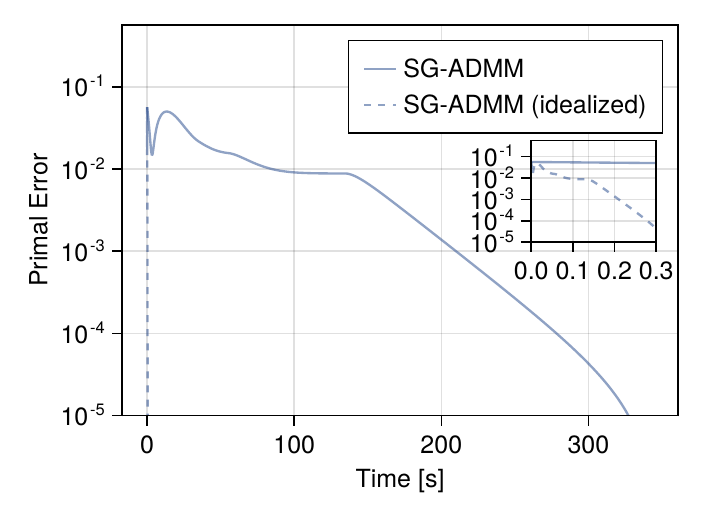}
        \caption{1000 sampled scenarios, under nonlinear unicycle dynamics \cite{laine2021computation}}\label{subfig:size 1000}
    \end{subfigure}
    \caption{Comparison of the CPU time under different numbers of sampled scenarios. The solid blue curves represent the implementation of Scenario-Game ADMM which solves step \ref{alg:parallel solving each scenario} of Algorithm~\ref{alg:scenario-admm} sequentially, i.e., one scenario by one scenario. The dashed blue curves represent the expected computation time when we implement step \ref{alg:parallel solving each scenario} of Algorithm~\ref{alg:scenario-admm} in parallel. This expected computation time is derived by dividing the computation time of the blue solid curves by the number of scenarios. In both cases, Scenario-Game ADMM converges faster than ACVI. For each sampled scenario, we have 20 dimensional decision variables, and 35 constraints. When the number of sampled scenarios is 1000, there are $1000\times 35=35000$ constraints. ACVI fails to compile due to the scale of problem. With 1000 samples, our algorithm converges even when we replace the linear dynamics in \eqref{eq: running example} with the nonlinear unicycle dynamics in \cite{laine2021computation}, as shown in Fig.\ref{subfig:size 1000}.    }
    \label{fig:comparison}
    \vspace{-0.5cm}
\end{figure}
\begin{theorem}\label{thm:ADMM convergence}
Under Assumption~\ref{assumption: ADMM convergence}, let $\zvar^* = (\auxs^*,\vars^*,\lms^*)$ be an optimal solution of \cref{eq:VI optimality condition}. Define $V(k):=(1/\rho)\|\lms(k)-\lms^*\|^2 + \rho\|M(\vars(k)-\vars^*)\|^2$. We have
\begin{equation}
    \begin{aligned}
        V(k+1) \le V(k) - \rho\|\auxs(k+1)-M \vars(k)\|^2\,,
    \end{aligned}
\end{equation}
and $\lim_{k\to \infty}V(k)=0$.
\end{theorem}
\begin{proof}
    The proof can be found in the Appendix.
\end{proof}


Theorem~\ref{thm:ADMM convergence} establishes the asymptotic convergence of Algorithm~\ref{assumption: ADMM convergence}, by showing that $V(k) \ra 0$ as $k \ra \infty$; thus, for any convergence tolerance $\epsilon > 0$, there exists some sufficiently large $k > 0$ such that $\|\auxs(k+1)-M\vars(k)\|^2\le \epsilon$. When the players' objectives satisfy the following assumption, we can strengthen the convergence result in Theorem~\ref{thm:strong ADMM convergence}.

\begin{assumption}[\cite{pavel2019distributed}]\label{assumption:strong ADMM convergence}
    For each player $i\in[N]$, the objective $f_i(\vars;\params)$ is differentiable. 
    The function $F(\auxs)=[\nabla_{\aux_i^j}f_i(\auxs^j;\params^j)]_{i=1,j=1}^{N,S}$ is $L$-Lipschitz continuous and is an $m$-strongly monotone operator, i.e., $(\auxs - \tilde{\auxs})^\top (F(\auxs) - F(\tilde{\auxs}))\ge m\|\auxs - \tilde{\auxs}\|_2^2, \forall \auxs,\tilde{\auxs}\in\mathbb{R}^{S\varsdim}$. {\color{black}The operator $\bar{F}(\auxs):=F(\auxs)-m\auxs$ is $1/(L-m)$-cocoercive, with $L>m$.} 
\end{assumption}

\begin{theorem}\label{thm:strong ADMM convergence}
    Under Assumptions~\ref{assumption: ADMM convergence} and \ref{assumption:strong ADMM convergence}, let $\zvar^*=(\auxs^*,\vars^*,\lms^*)$ be an optimal solution of \eqref{eq:VI optimality condition}. Define $V(k)$ as in Theorem~\ref{thm:ADMM convergence}. At the $k$-th iteration, we have:
    \begin{enumerate}
        \item If there is no binding constraint at $\auxs(k)$, i.e., $h_i(\vars(k);\params^j)<0$, $\forall i\in[N]$, $\forall j\in [S]$, then:
        \begin{equation*}
            V(k) \le \left(1-\frac{1}{2 \kappa_ f^{0.5+|\epsilon|}} \right) V(k-1)\,
        \end{equation*}
        where $\kappa_f = L/m$ and $\epsilon = \log_{\kappa_f}(\rho/\sqrt{mL})$;
        \item Otherwise, $V(k) \le V(k-1)-\rho\|\auxs(k) - M\vars(k-1)\|^2$.
    \end{enumerate}
\end{theorem}
\begin{proof}
    The proof can be found in the Appendix.
\end{proof}


%% file: 6_Results.tex
\section{Experiments}
\label{sec: Results}

In this section, we continue the running example \eqref{eq: running example}. We characterize the sample complexity and the empirical performance of the Scenario-Game ADMM. The details of the experiment parameters are included in the Appendix. 

By Proposition~\ref{prop: sample complexity}, if the sample size is $S=1000$, then for each player $i\in\{1,2\}$, $\mathbb{P}(\|\frac{1}{S} \sum_{j=1}^Sf_i(\vars;\params^j) - \mathbb{E}[f_i(\vars)]\|\le 0.5)\ge 1-4.0\times 10^{-3}$, and $\mathbb{P}(\mathbb{P}(h_i(\vars;\params)\le 0)\ge 0.95)\ge 1-2.9\times10^{-7}$. Therefore, by having $1000$ sampled scenarios, we are able to obtain a reasonable approximation \eqref{eqn:scenario-game} of the stochastic game problem \eqref{eq: running example}. 

We proceed to apply Scenario-Game ADMM to solve the sample-approximated game problem \eqref{eqn:scenario-game}. We first validate the convergence of Scenario-Game ADMM in Fig.~\ref{fig:scenario-game ADMM convergence}. As proven in Theorem~\ref{thm:ADMM convergence}, the Lyapunov function decays monotonically in Fig.~\ref{subfig:lyapunov}. Note that the primal residual $\rho\|M(\vars(k)-\vars^*)\|^2$ may still oscillate due to the existence of binding constraints, as shown in Theorem~\ref{thm:strong ADMM convergence} and Fig.~\ref{subfig:primal residual}.

We then compare the performance of Scenario-Game ADMM with the baseline method. Since prior works \cite{liang2017distributed,pavel2019distributed} do not consider coupled nonlinear constraints among players, we compare Scenario-Game ADMM with the state-of-the-art ADMM-based constrained variational inequality solver (ACVI) \cite{yang2022solving}. As shown in Fig.~\ref{fig:comparison}, Scenario-Game ADMM converges faster than ACVI across different scenario sizes. In particular, when we have 1000 sampled scenarios, Scenario-Game ADMM converges, but ACVI fails to compile due to the scale of the problem, 
where we have $35000$ coupled inequality constraints in total. 
This experiment suggests that Scenario-Game ADMM can solve game problems with a large number of constraints within a reasonable amount of time. 


As an additional ablation, we also compare our method's computation time to the centralized PATH solver that our method uses at the inner loop 
\cite{dirkse1995path}; \cf appendix.
While PATH is competitive, in particular for small-scale problems, we observe that the parallelized version of our method is still more than 2x faster for scenario sizes $S \in [10, 100]$.
Finally, as with ACVI, the scenario-number-dependent compilation overhead of this centralized approach precludes application to larger problems. 

%% file: 7_Conclusion_Future_Work.tex
\section{Conclusion and Future Work}
\label{sec: Conclusion and Future Work}
In this work, we introduced a new sample-based approximation for stochastic games. We characterized the sample complexity and the feasibility guarantees of this approximation scheme. We proposed a decentralized ADMM solver and characterized its convergence. We empirically validated the performance of this algorithm in a stochastic game with a large number of sampled constraints. Future work should extend our results on sample complexity and analyze how well equilibria of the scenario game approximate solutions to the original chance-constrained stochastic game.

\section*{Acknowledgements}
This work is supported by the DARPA Assured Autonomy and ANSR programs, the
NASA ULI program in Safe Aviation Autonomy, and the ONR Basic Research Challenge in Multibody Control Systems. This work is also supported by the National Science Foundation under Grant Nos. 2211548 and 1652113, and the Army Research Laboratory under Cooperative Agreement Numbers W911NF-23-2-0011 and W911NF-20-1-0140.

%% file: A_Appendix.tex
\appendix
\label{sec: Appendix}
\noindent \textbf{Experiment Details.} 
The random cost matrix $Q_i^\params$ is parameterized as $Q_i^\params =I_4 + P_i^{\params\top}P_i^\params$, where each entry of $P_i^\params\in\mathbb{R}^{4\times4}$ is uniformly sampled from $[0,1]$. Each entry of the constraint parameter $b_i^\params$ is sampled from $[0,0.01]$. $\xi_1^x(0)$ and $\xi_1^y(0)$ are uniformly sampled from $[-0.15, 0.0 ]$. $\xi_2^x(0)$ and $\xi_2^y(0)$ are uniformally sampled from $[0.0, 0.15]$. Both players have zero initial velocity. We can verify that an upper bound of the cost function in \eqref{eq: running example} for all feasible control inputs is $D=3$. The decision variable of each player is its control input. For each sampled scenario, the total dimension of decision variables is $T\times N\times q=20$, where $T=5$ is the horizon, $N=2$ is the number of players, and $q=2$ is the control input dimension of one player at each time instance $t\in [T]$. We pick $\rho = 5$. We use PATH \cite{dirkse1995path} to compute the inner MCP problems in Scenario-Game ADMM and ACVI. For ACVI, we adopt the best parameters we found: the log-penalty coefficients $\mu_t$ are defined as $\mu_t = (\frac{1}{2})^t\mu_0$, where $t$ is the outer iteration number of the interior point method \cite{yang2022solving} and $\mu_0 = 10^{-4}$.

%% file: A_Appendix_proof.tex
\noindent \textbf{Proofs.} Before we present the proof of Proposition~\ref{prop: sample complexity}, we first introduce the following lemmas.
\begin{lemma}[Thm. 3.26, \cite{wainwright2019high}]\label{lem: sample complexity of objective}
\looseness=-1
    Let $\{\theta^j\}_{j=1}^S$ be i.i.d. samples from $p_\theta$. Suppose $\exists D$, s.t. $\sup_{\theta\in \Theta,\vars\in\mathcal{X}}\|f(\vars;\theta)\|_2\le D<\infty.$ Then, $\mathbb{P}( \|\sup_{\vars\in \mathcal{X}} \frac{1}{S}\sum_{j=1}^S f(\vars;\theta^j) - \mathbb{E}_\params[f(\vars;\params)]\|_2\ge \tilde{\epsilon} ) \le 2e^{(-\frac{S \tilde{\epsilon}^2 }{4D^2})}$.
\end{lemma}
\begin{lemma}[\cite{campi2008exact}]\label{lem: sample complexity of constraint}
    Let $\{\params^j\}_{j=1}^S$ be a set of i.i.d. samples of the random variable $\params$. For all $\vars \in \mathbb{R}^{Nn}$, we have $ \mathbb{P}\left(\mathbb{P}\left( h(\vars;\params)\le0 \right)\ge \epsilon\right)\le \sum_{\ell = 0}^{Nn-1}\binom{S}{\ell}\epsilon^{\ell}(1-\epsilon)^{S - \ell}$.
\end{lemma}

\begin{proof}[Proof of Proposition~\ref{prop: sample complexity}]
    By Assumption~\ref{assumption: ADMM convergence}, $\sup_{\params\in\Theta,\vars\in \mathcal{X}}f_i(\vars;\theta)\le D$, for some finite $D\in\mathbb{R}$. Let $h(\vars;\params):=[h_i(\vars;\params)]_{i=1}^N$. By Lemmas~\ref{lem: sample complexity of objective} and \ref{lem: sample complexity of constraint} and the union bound, $\mathbb{P}( \sup_{\vars\in\mathcal{X}}\|\frac{1}{S}\sum_{j=1}^S f_i(\vars;\params^j)-\mathbb{E}_\params[f_i(\vars;\params)]\|\le \tilde{\epsilon}\textrm{ and }   \mathbb{P}\left(h(\vars;\theta)\le 0\right)\ge 1-\epsilon,\forall i\in[N]  ) \ge 1 - 2Ne^{(-\frac{S\tilde{\epsilon}^2}{4D^2})} - \sum_{\ell = 0}^{Nn-1}\binom{S}{\ell}\epsilon^{\ell}(1-\epsilon)^{S - \ell}$.
\end{proof}
\begin{proof}[Proof of Proposition~\ref{prop: improved sample complexity}]
    Under the independent constraint assumption, we have $\mathbb{P}(\mathbb{P}(h_i(\var_i;\params)\le 0)\ge \epsilon) \le \sum_{\ell=0}^{n-1}\binom{S}{\ell} \epsilon^\ell (1-\epsilon)^{S-\ell}$.
    Then, by Lemma~\ref{lem: sample complexity of objective} and the union bound, we have $\mathbb{P}( \sup_{\vars\in\mathcal{X}}\|\frac{1}{S}\sum_{j=1}^S f_i(\vars;\params^j)-\mathbb{E}_\params[f_i(\vars;\params)]\|\le \tilde{\epsilon}\textrm{ and } \mathbb{P}\left(h_i(\vars;\theta)\le 0\right)\ge 1-\epsilon,\forall i\in[N]  ) \ge 1 - 2Ne^{(-\frac{S\tilde{\epsilon}^2}{4D^2})} - N\sum_{\ell = 0}^{n-1}\binom{S}{\ell}\epsilon^{\ell}(1-\epsilon)^{S - \ell}$.
\end{proof}

\begin{proof}[Proof of Lemma \ref{lem: consensus constraint residual}]
From Algorithm~\ref{alg:scenario-admm}, we have $        (\auxs-\auxs(k+1))^\top (F(\auxs(k+1)) +H(\auxs(k+1))+ \lms(k))  \ge 0,\forall \auxs$, $ 
        \vars(k+1) = M^\dag (\auxs(k+1) + (1/\rho)\lms(k))$, and $
        \lms(k+1) = \lms(k) + \rho ( \auxs(k+1) - M \vars(k+1))
$.
Since $M$ has full column rank, we have the null space of $M$ is $\{0\}$. Also, by definition, $\sum_{j=1}^S \lambda_i^j(k+1) = \sum_{j=1}^S \lambda_i^j(k) + \rho (\sum_{j=1}^S w_i^j(k+1)- Mx_i(k+1))=\sum_{j=1}^S \lambda_i^j(k)-\sum_{j=1}^S \lambda_i^j(k)=0$, and therefore $M^\dag \lms(k)=0$. Thus, $\vars(k+1) = M^\dag \auxs(k+1) = \vars(k)$. This implies that $\auxs(k+1)-M\vars(k+1)=0$, and $\lms(k+1) = \lms(k)$. $(\auxs(k+1),\vars(k+1),\lms(k+1))$ satisfies the optimality condition \eqref{eq:VI optimality condition}.
\end{proof}


\begin{lemma}\label{lem:key lemma for ADMM convergence}
    Let $(\vars^*,\auxs^*,\lms^*)$  be an optimal solution to \eqref{eqn: optimal condition for variational inequality}, it holds $        (1/\rho)(\lms(k+1) - \lms^*)^\top (\lms(k+1) - \lms(k))  \le \rho (\auxs(k+1) - \auxs^*)^\top(M\vars(k)-M\vars(k+1) ) $.
\end{lemma}

\begin{proof}
    From the optimality condition~\eqref{eq:VI optimality condition}, we have:
    {\small
    \begin{equation}\label{eq:w^* optimality condition}
        (\auxs(k+1) - \auxs^*)^\top (F(\auxs^*)+H(\auxs^*) + \lms^*)\ge 0
    \end{equation}
    }
    By the optimality condition at the $k$-th iteration, we have $(\auxs^* - \auxs(k+1))^\top (F(\auxs(k+1))+H(\auxs(k+1)) + \lms(k) + \rho (\auxs(k+1) - M\vars(k)))\ge 0$.
    Substituting $\lms(k+1) = \lms(k)+\rho(\auxs(k+1)-M\vars(k+1))$, we derive:
    {\small
    \begin{equation}\label{eq:w(k+1) optimality condition}
        \begin{aligned}
            &(\auxs^*-\auxs(k+1))^\top (F(\auxs(k+1))+H(\auxs(k+1))\\ &+\lms(k+1) + \rho(M(\vars(k+1) - M\vars(k))) )\ge 0
        \end{aligned}
    \end{equation}
    }
    Adding \eqref{eq:w^* optimality condition} and \eqref{eq:w(k+1) optimality condition}, and using monotonicity, we have:
    {\small
    \begin{equation}\label{eq:w monotonicity bound}
    \begin{aligned}
        &(\auxs(k+1)-\auxs^*)^\top (\lms(k+1) - \lms^*) \\ &\le \rho(\auxs(k+1)-\auxs^*)^\top M(\vars(k)-\vars(k+1))
    \end{aligned}
    \end{equation}
    }
    Similarly, by the optimality of $\vars^*$ and $\vars(k+1)$, we have $(\vars(k+1)-\vars^*)^\top ( - M^\top\lms^*)\ge 0$ and $(\vars^* - \vars(k+1))^\top (-M^\top\lms(k+1))\ge0$. Adding these two inequalities:
    {\small
    \begin{equation}\label{eq:x monotonicity bound}
        \begin{aligned}
            (M\vars^*-M\vars(k+1))^\top  (\lms(k+1)-\lms^*)\le 0
        \end{aligned}
    \end{equation}
    }
    Adding \eqref{eq:w monotonicity bound} and \eqref{eq:x monotonicity bound}, and using $\auxs^*-M\vars^*=0$ and $\auxs(k+1)-M\vars(k+1)=\frac{1}{\rho}(\lms(k+1)-\lms(k))$, we have $\frac{1}{\rho}(\lms(k+1)-\lms(k))^\top (\lms(k+1)-\lms^*) \le \rho(\auxs(k+1)-\auxs^*)^\top (M\vars(k) - M\vars(k+1))$.
\end{proof}

\begin{proof}[Proof of Theorem~\ref{thm:ADMM convergence}]
    Observe that:
    {\small
    \begin{equation}\label{eq:ADMM convergence proof main quadratic expansion}
    \begin{aligned}
        &(1/\rho) \|\lms(k+1) - \lms^*\|^2 + \rho\|M(\vars(k+1) - \vars^*)\|^2\\
        &=(1/\rho) \|\lms(k) - \lms^*\|^2 + \rho\|M(\vars(k)-\vars^*)\|^2\\
        &\ \ -((1/\rho)\|\lms(k+1) - \lms(k)\|^2 + \rho \|M(\vars(k+1)-\vars(k))\|^2)\\
        &\ \  + (2/\rho)(\lms^*-\lms(k+1))^\top (\lms(k)-\lms(k+1))\\
        &\ \  + 2\rho (M\vars^* - M\vars(k+1))^\top (M\vars(k) - M\vars(k+1))
    \end{aligned}
    \end{equation}
    }
    The last two terms can be bounded as:
    {\small
    \begin{equation}\label{eq:ADMM convergence proof last two terms bound}
        \begin{aligned}
            &(2/\rho)(\lms^*-\lms(k+1))^\top (\lms(k)-\lms(k+1))\\
        &\ \  + 2\rho (M\vars^* - M\vars(k+1))^\top (M\vars(k) - M\vars(k+1))\\
        &\le  2\rho(\auxs(k+1)-\auxs^*)^\top M(\vars(k)-\vars(k+1))\\
        &\ +2\rho(M\vars^* - M\vars(k+1))^\top (M\vars(k)-M\vars(k+1))\\
        &=2\rho (\auxs(k+1)+M\vars(k+1))^\top(M\vars(k)-M\vars(k+1))\\
        &=-2(\lms(k)-\lms(k+1))^\top(M(\vars(k)-\vars(k+1)))
        \end{aligned}
    \end{equation}
    }
    where the first inequality follows from Lemma~\ref{lem:key lemma for ADMM convergence}, and the first equality is derived by substituting $\auxs^*-M\vars^*=0$. The last equality holds true because of the update rule of $\lms(k+1)$. 

    
    From \eqref{eq:ADMM convergence proof main quadratic expansion} and \eqref{eq:ADMM convergence proof last two terms bound}, we have $(1/\rho)\|\lms(k+1)-\lms^*\|^2 + \rho \|M(\vars(k+1)-\vars^*)\|^2
            \le 
            (1/\rho)\|\lms(k)-\lms^*\|^2+\rho\|M(\vars(k)-\vars^*)\|^2
             -\rho\|\auxs(k+1)-M\vars(k)\|^2$.
\end{proof}

Before we present the proof of Theorem~\ref{thm:strong ADMM convergence}, we first introduce a few preliminaries. Define $\hat{f}_i:=\frac{1}{\rho}f_i$ and $\hat{g}:=\mathbb{I}_{\textrm{im}\ M}$, where $\mathbb{I}_{\textrm{im}\ M}$ is the $\{0,\infty\}$-indicator function of the image of $M$. Additionally, we define $s(k):=M\vars(k)$, $u(k) := \lambda(k)/\rho$. Let $\beta(k):= [\nabla_{\var_i} \hat{f}_i(\auxs^j(k);\theta^j)]_{i=1,j=1}^{N,S}$ and $\gamma(k) := [\partial_{\var_i}\hat{g}(\vars)]_{i=1,j=1}^{N,S}$. 
\begin{lemma}\label{lem:strong monotone gradient}
    Under Assumption~\ref{assumption:strong ADMM convergence}, let $\auxs,\tilde{\auxs}\in\mathbb{R}^{SNn} $, $E = [\nabla_{\aux_i^j}f_i(\auxs^j;\params^j)]_{i=1,j=1}^{N,S}$ and $\tilde{E}=[\nabla_{\aux_i^j}f_i(\tilde{\auxs}^j;\params^j)]_{i=1,j=1}^{N,S}$. We have $\left[\begin{smallmatrix}
        \auxs-\tilde{\auxs} \\ E-\tilde{E}
    \end{smallmatrix}\right]^\top \left[\begin{smallmatrix}
        -2mL & m+L \\ m+L & -2
    \end{smallmatrix}\right]\otimes I_{SNn}\left[\begin{smallmatrix}
            \auxs - \tilde{\auxs} \\ E - \tilde{E}
    \end{smallmatrix}\right]\ge 0$.
\end{lemma}
\begin{proof}
    \color{black}By Assumption~\ref{assumption:strong ADMM convergence}, we have $(m+L)(\auxs - \tilde{\auxs})^\top (E-\tilde{E})\ge mL\|\auxs - \tilde{\auxs}\|_2^2 + \|E-\tilde{E}\|_2^2$.
    We complete the proof by putting it in matrix form.
\end{proof}

\begin{lemma}\label{lem:ADMM dynamics}
    Suppose there is no binding constraint at $\auxs(k+1)$. Let $\eta(k):=[s(k), u(k)]$, $v(k):= [\beta(k+1), \gamma(k+1)]$, $y(k):=[\auxs(k+1), \beta(k+1)]$ and $z(k):=[s(k+1), \gamma(k+1)]$. We consider $\eta(k)$, $v(k)$ and $[y(k),z(k)]$ as the state, control input and output of a dynamical system. Define the following matrices, $\hat{A}:=\left[\begin{smallmatrix}1 & 0 \\ 0 & 0\end{smallmatrix}\right]$, $\hat{B}:=\left[\begin{smallmatrix}  
            -1 & -1 \\ 0 &-1
\end{smallmatrix}\right]$, $\hat{C}^1:=\left[\begin{smallmatrix}
    1 & -1 \\ 0 &0
\end{smallmatrix}\right]$, $\hat{D}^1:=\left[\begin{smallmatrix}
    -1 & 0 \\ 1 & 0
\end{smallmatrix}\right]$, $\hat{C}^2:=\left[\begin{smallmatrix}
    1 & 0 \\ 0 & 0
\end{smallmatrix}\right]$, and $\hat{D}^2:=\left[\begin{smallmatrix}
    -1 & -1 \\ 0 & 1
\end{smallmatrix}\right]$.
    Then, we have the dynamics $\left[\begin{smallmatrix}
            \eta(k+1) \\ y(k) \\ z(k)
        \end{smallmatrix}\right] = \left[\begin{smallmatrix}
            \hat{A} &  \hat{B} \\ \hat{C}^1 & \hat{D}^1  \\ \hat{C}^2  & \hat{D}^2 
        \end{smallmatrix}\right]\left[\begin{smallmatrix}
            \eta(k) \\ v(k)
        \end{smallmatrix}\right]$.
\end{lemma}
\begin{proof}
    By the KKT condition, we have $\nabla_i f_i (\auxs^j(k+1);\theta^j) + \lambda_i^j(k)+ \rho(\auxs^j(k+1) - Mx_i(k)) = 0$, which is equivalent to 
    {\small
    \begin{equation}\label{eq:dynamics for w}
        \beta(k+1)  + u(k) + \auxs(k+1) - s(k) = 0
    \end{equation}
    }
    Subsequently, when we minimize $\vars$ with $\auxs(k+1)$ and $\lambda(k)$ fixed, we have the problem of minimizing $\vars$ is equivalent to, for each $i\in[N]$, $\min_{s_i} \partial_i \hat{g}(s_i) +u_i(k)^\top (\auxs_i(k+1)-s_i) +  \frac{\rho}{2}\|\auxs_i(k+1) - s_i\|^2$:
    which has the optimality condition:
    {\small
    \begin{equation}\label{eq:dynamics for x}
        \gamma(k+1)+s(k+1) - \auxs(k+1) - u(k) = 0.
    \end{equation}
    }
    Finally, from the update rule of the Lagrange multiplier, we have $\lambda(k+1) = \lambda(k) + \rho (\auxs(k+1) - B\vars(k+1))$, and this implies:
    {\small
    \begin{equation}\label{eq:dynamics for lambda}
        u(k+1) = u(k) + \auxs(k+1) - s(k+1) = \gamma(k+1)
    \end{equation}
    }
    where the last equality follows by substituting \eqref{eq:dynamics for x}. We complete the proof by rearranging terms in \eqref{eq:dynamics for w}-\eqref{eq:dynamics for lambda}.
\end{proof}
\begin{proof}[Proof of Theorem~\ref{thm:strong ADMM convergence}]
    The second part has been shown in Theorem~\ref{thm:ADMM convergence}, we only need to prove the first part. Note that the gradient of $\hat{f}$ is $\frac{\rho}{(mL)^{1/2}}\kappa_f^{-1/2}$-Strongly monotone and $\frac{\rho}{(mL)^{1/2}}\kappa_f^{1/2}$-Lipschitz, and $M$ is full column rank. 
    We can extend Theorem 6 \cite{nishihara2015general} to variational inequality problem by using Lemma~\ref{lem:strong monotone gradient}, and Lemma~\ref{lem:ADMM dynamics}. Then, by Theorem~7 \cite{nishihara2015general}, we have
    $V(k) \le  (1-1/{(2 \kappa_f^{0.5+|\epsilon|})}) V(k-1)$,
    where $\epsilon = \log_{\kappa_f}(\rho/{\sqrt{mL}})$.
\end{proof}